\documentclass[usenatbib]{mn2e}
\usepackage{amsmath}
\usepackage{graphicx}
\usepackage{hyperref}
\usepackage[T1]{fontenc}
\newcommand{\mesa}{\texttt{MESA star}~}
\hypersetup{colorlinks,linkcolor=blue, citecolor=blue}

\newcounter{species} 
\def\ion#1#2{\setcounter{species}{#2}#1$\;${\scriptsize\Roman{species}}\relax}
\begin{document}
\title[Emission Lines from Tidally Disrupted HB Stars]{Luminous [O III] and [N II] from Tidally Disrupted Horizontal Branch Stars}
\author[Clausen et al.]{Drew Clausen,$^{1}$\thanks{E-mail:dclausen@astro.psu.edu} Steinn Sigur\dh sson,$^{1}$ Michael Eracleous,$^{1}$ Jimmy A. Irwin$^{2}$\\
$^{1}$Department of Astronomy \& Astrophysics, The Pennsylvania State University, 525 Davey Lab, University Park, PA 16802\\
$^{2}$Department of Physics and Astronomy, University of Alabama, Box 870324, Tuscaloosa, AL 35487, USA}
\newcommand{\apj}{ApJ}
\newcommand{\apjs}{ApJS}
\newcommand{\apjl}{ApJL}
\newcommand{\mnras}{MNRAS}
\newcommand{\aap}{A\&A}
\newcommand{\araa}{ARA\&A}
\newcommand{\nat}{Nat}
\newcommand{\aj}{AJ}
\newcommand{\pasp}{PASP}
\newcommand{\aaps}{A\&AS}
\newcommand{\nii}{[\ion{N}{2}] $\lambda 6583$}
\newcommand{\oiii}{[\ion{O}{3}] $\lambda 5007$}
\newcommand{\jnii}{[\ion{N}{2}]}
\newcommand{\joiii}{[\ion{O}{3}]}
\newcommand{\src}{CXOJ033831.8 - 352604}
\maketitle

\begin{abstract}
We model the emission lines generated in the photoionised debris of a tidally disrupted horizontal branch star.  We find that at late times, the brightest optical emission lines are [\ion{N}{2}] $\lambda\lambda 6548,6583$ and [\ion{O}{3}] $\lambda\lambda 4959,5007$.  Models of a red clump horizontal branch star undergoing mild disruption by a massive (50 -- 100 $M_{\sun}$) black hole yield an emission line spectrum that is in good agreement with that observed in the NGC 1399 globular cluster hosting the ultraluminous X-ray source CXOJ033831.8 - 352604.  We make predictions for the UV emission line spectrum that can verify the tidal disruption scenario and constrain the mass of the BH.           
\end{abstract}


\section{Introduction}   
With the exquisite angular resolution of {\em Chandra}, it is possible to resolve low mass X-ray binaries (LMXBs) in nearby galaxies and to associate these LMXBs with globular clusters \citep[GCs; e.g.,][]{Sarazin:2000}.  This capability has enabled the discovery of several black hole (BH) candidates in GCs outside of the Milky Way.  \citet{Maccarone:2007} presented the first strong candidate for a BH in a GC, an ultraluminous X-ray (ULX) source in the globular cluster RZ 2109 associated with the elliptical galaxy NGC 4472.  The object's X-ray luminosity of $L_{\rm X} =4\times 10^{39}~{\rm erg\;s^{-1}}$ is too large to be explained by accretion onto a neutron star (NS), and the source's strong variability rules out the possibility of a superposition of multiple NSs.  Subsequently, three additional ULXs have been identified as GC BH candidates on the basis of their strong X-ray variability; one in NGC 1399 \citep{Shih:2010}, one in NGC 3379 \citep{Brassington:2010}, and a second candidate in NGC 4472 \citep{Maccarone:2011a}.  Variable X-ray sources with luminosities above the Eddington limit of a NS offer the least ambiguous evidence of a BH accretor; however, \citet{Barnard:2011} argued for BH accretors in three LMXBs located in M31 GCs that exhibit low/hard-state spectra at luminosities near a NS's Eddington limit, and in a high luminosity recurrent transient in a fourth M31 GC. 

\citet[][hereafter I10]{Irwin:2010} discussed an additional BH candidate in an NGC 1399 GC; however,  the nature of this source is not as clear-cut as that of the other candidates.   The object's X-ray luminosity, $L_{\rm X} \sim 2 \times 10^{39}~{\rm erg~s^{-1}}$, suggests a BH accretor, but the 35\% decline in the luminosity between 2000 and 2008 does not convincingly rule out a superposition of sources.  The X-ray spectrum can be fitted with a power law of photon index $\Gamma = 2.5$, which is much softer than the slope of the power law used to fit the sum of low $L_{\rm X}$ LMXBs ($\Gamma \sim  1.6$). This is also slightly softer than the spectra of other sources with $L_{\rm X} \sim 10^{39}\;{\rm erg~s^{-1}}$, which might indicate the presence of a more massive BH.  Adding to the intrigue, optical spectroscopy of the NGC 1399 GC revealed bright  [\ion{N}{2}] $\lambda\lambda 6548,6583$ and [\ion{O}{3}] $\lambda\lambda 4959,5007$ emission lines with luminosities $\sim$$10^{36}~{\rm erg~s^{-1}}$, but no Balmer emission lines.  This is not the only GC BH with optical emission lines, \citet{Zepf:2008} found bright, broad \oiii~emission lines without accompanying Balmer lines in the optical spectrum of RZ 2109.  I10 argued that these two sources could be evidence of the tidal disruption or detonation of a white dwarf by an intermediate mass BH, based on the lack of hydrogen emission lines in the spectra.  

A BH in the core of a galaxy or a GC can tidally disrupt stars that pass too closely, and then accrete a portion of the debris \citep{Hills:1975,Frank:1976,Sigurdsson:1997,Baumgardt:2004,Ramirez-Ruiz:2009}.   For stars initially on parabolic orbits with the BH, roughly half of the mass of the disrupted star will become bound to the BH, and the other half will remain unbound and stream back out into the cluster \citep{Lacy:1982,Rees:1988}.  \citet{Roos:1992} first considered the emission lines produced when the unbound debris is photoionised by the UV/X-ray emission generated when the bound portion accretes onto the BH.  Detailed studies of the emission lines produced when a main-sequence star is tidally disrupted by a supermassive BH have been made using numerical  \citep{Bogdanovic:2004} and analytic \citep{Strubbe:2009,Strubbe:2011} models for the dynamical evolution of the debris.  \citet{Clausen:2011} considered the case of a white dwarf being tidally disrupted by an intermediate mass BH and found that the emission lines observed in the two BH candidates described above were not consistent with those expected from a tidally disrupted white dwarf.  I10 reported that the \jnii~and \joiii~doublets observed in the NGC 1399 GC have roughly the same luminosity, but the models predict that the \jnii~doublet should be two orders of magnitude fainter than the \joiii~doublet.  The white dwarf tidal disruption models predict that the \oiii~luminosity will reach a peak value consistent with that observed in RZ 2109 a few years after tidal disruption.  However, the X-ray source in RZ 2109 was detected with {\em ROSAT} in 1994 \citep{Colbert:2002}, which indicates that the observations reported in  \citet{Zepf:2008} were made more than a decade after the BH began accreting.      

Other models have been proposed to explain the emission lines observed in GCs hosting BH candidates.  \citet{Maccarone:2010} found that the X-ray emission from RZ 2109 had varied by a factor of 20 over a period of six years, a timescale far too short to be consistent with a tidal disruption, and suggested that the source is a hierarchical triple system with an inner BH-white dwarf binary.  Such a system is in line with the work of \citet{Ivanova:2010}, who suggested that a triple system may be necessary to form a BH-white dwarf binary. \citet{Ripamonti:2012} showed that nova ejecta that have been photoionised by a ULX could produce bright, broad \joiii~emission lines similar to those observed in RZ 2109.  \citet{Porter:2010} considered an accretion disc origin for the emission lines in RZ 2109 and the GC in NGC 1399.  \citet{Maccarone:2011} argued that the lines observed in the NGC 1399 GC were formed in the photoionised winds of R Corona Borealis (RCB) stars.  Here we propose that the X-ray continuum and optical emission lines observed in the NGC 1399 are the result of a tidal disruption of a horizontal branch (HB) star.  In \autoref{models} we describe the ingredients of our models, in \autoref{results} we describe the results of our calculations, and in \autoref{discussion} we compare our model with competing ones and discuss an observational test to distinguish between them. 
\begin{figure}
	\centering
	\includegraphics[width=0.45\textwidth]{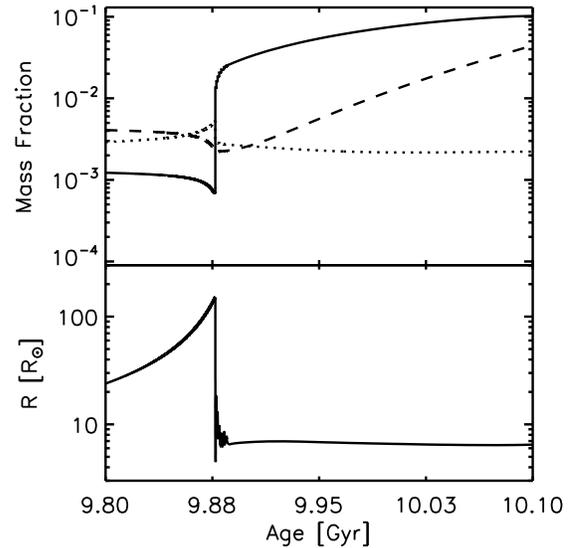}
	\caption{Mass fractions ({\it top}) of carbon ({\it solid}), nitrogen ({\it dotted}), and oxygen ({\it dashed}) vs. time and radius ({\it bottom}) vs. time from the \mesa~model of a 1 $M_{\sun}$ star with $Z = Z_{\sun}/2$.  \label{mesa}}
\end{figure}

\section{Models}
\label{models}
We modelled the emission-line spectrum produced when a HB star on an unbound orbit is disrupted by a BH, using the method outlined in \citet{Clausen:2011}, which is an adaptation of the procedure used by \citet{Strubbe:2009}.  Our approach combines analytic prescriptions for the dynamical evolution of the debris with {\em Cloudy} \citep{Ferland:1998} photoionisation models to compute the luminosities and line-profiles of the emission lines produced in the unbound portion of the debris.  These models require as inputs the mass of the disrupted star $M_{\star}$, the radius of the disrupted star $R_{\star}$, the mass of the BH $M_{\rm BH}$, the pericentre distance of the orbit $R_{\rm P}$, and the composition of the disrupted star.  We determined the properties of the HB stars used in our models using the stellar evolution code \mesa \citep{Paxton:2011}. 

\begin{figure}
	\centering
	\includegraphics[width=0.45\textwidth]{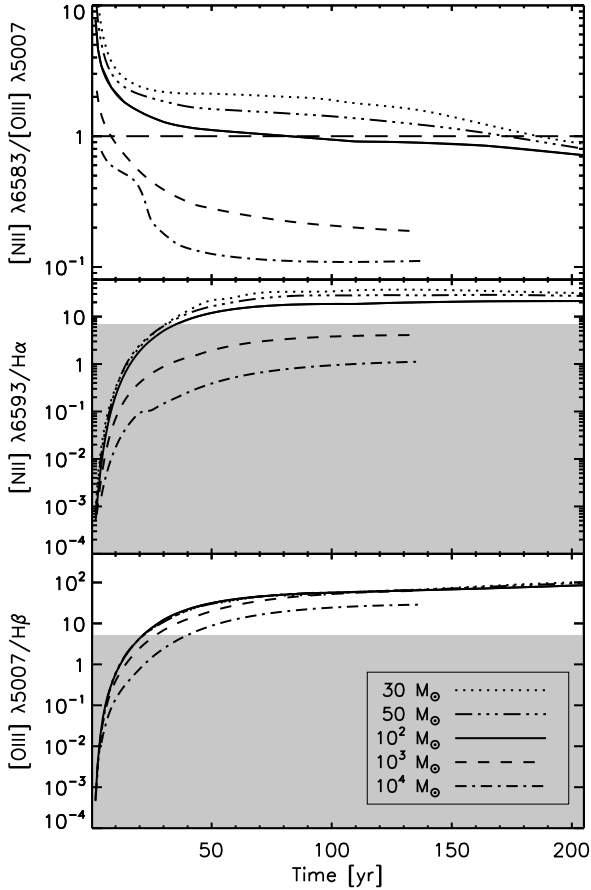}
	\caption{Emission-line luminosity ratios over time from several tidal disruption models with varying $M_{BH}$.  The long-dashed line in the top panel shows the constraint on the \jnii/\joiii~ratio measured by I10. Models that fall near this line are consistent with observations.  In the lower two panels, the shaded region indicates line ratios ruled out by the observations of I10, only models that reach into the unshaded region are consistent with observations.   For each model shown here, we considered a red clump star on an orbit with $R_{P} = R_{T}$.  \label{ratios}}
\end{figure}
\begin{figure}
	\centering
	\includegraphics[width=0.45\textwidth]{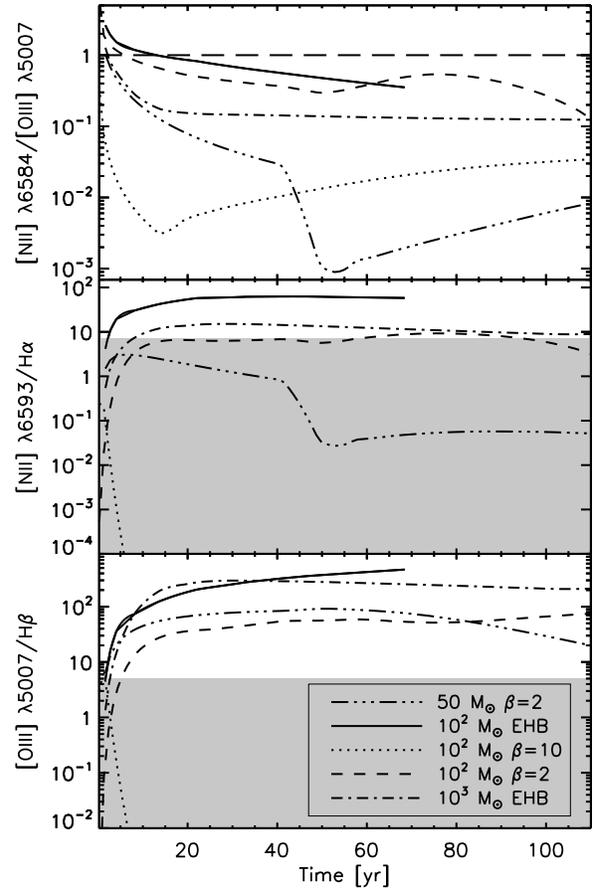}
	\caption{Emission-line luminosity ratios over time from several tidal disruption models with varying $M_{BH}$, tidal disruption parameter $\beta$, and stellar type.  The long-dashed line in the top panel shows the constraint on the \jnii/\joiii~ratio measured by I10. Models that fall near this line are consistent with observations.  In the lower two panels, the shaded region indicates line ratios ruled out by the observations of I10, only models that reach into the unshaded region are consistent with observations.\label{odd}}
\end{figure}

\subsection{Horizontal Branch Star Models}

A GC's horizontal branch morphology is strongly influenced by its metallicity, with the HBs in most metal rich GCs lying redward of the instability strip and those of most low metallicity clusters lying to the blue side.  However, it has been known for some time that metallicity alone cannot account for the range of HB morphologies observed, and that HB morphology is determined by metallicity and at least a second parameter  \citep[e.g.,][]{Sandage:1960,van-den-Bergh:1967}.  {  By the mid 1990s, a number of studies \citep[e.g.][]{Searle:1978, Lee:1994} had established cluster age as a likely solution
to the second parameter problem. Recent, independent studies by \citet{Dotter:2010} and \citet{Gratton:2010} have confirmed that the second parameter is the age of the GC.}  The GC hosting the X-ray source considered here is quite red \citep[$B-I=2.25$;][]{Kundu:2007}, and therefore likely has a HB consisting primarily of red clump stars, which is typical of high metallically GCs.  However, there is evidence that some metal rich GCs harbour extreme horizontal branch (EHB) and even blue hook stars \citep[e.g.][]{Dalessandro:2008, Rey:2007, Peacock:2010}, so we will consider the tidal disruption of both red clump and EHB stars.

Using \texttt{MESA star}, we evolved a $1~M_{\sun}$ star with $Z = Z_{\sun}/2$ to determine the mass, radius, and composition of the red clump star.  The star evolved to the horizontal branch after $\sim 9.9$ Gyr, at which point $M_{\star} = 0.67~M_{\sun}$ and $R_{\star} =6.2~R_{\sun}$.  Furthermore, CNO cycle burning had enriched the core of the star with nitrogen at the expense of carbon and oxygen.  Leading up to the HB phase, the total mass fraction of nitrogen rose and those of oxygen and carbon fell as the star's relatively carbon and oxygen rich envelope was shed.  The ratio of nitrogen to oxygen is highest right after helium ignition, at the start of the HB phase. We used the mass, radius, and composition of the HB star at this point as our fiducial model.  At this point the star was made up of 15\% H, 84\% He, 0.1\% C, 0.5\% N, and 0.2\% O, by mass, with all other elements present in their initial mass fraction.  The mass and radius of our adopted HB star is consistent with previous models \citep[e.g.,][]{Dorman:1992, Charbonnel:1996}.  Furthermore, measurements of the surface abundances in HB stars have revealed a nitrogen enhancement, but the surface abundance of nitrogen, which depends on the stars mixing history, is highly variable from star to star \citep[e.g.,][]{Behr:1999,Gratton:2000,Tautvaisiene:2001}.    

We found $M_{\star} = 0.49~M_{\sun}$ and $R_{\star} = 0.12~R_{\sun}$ for an EHB star by running a \mesa model of a $1~M_{\sun}$ star with $Z= 0.1~Z_{\sun}$.  This combination of $M_{\star}$ and $R_{\star}$ are in good agreement with the asteroseismology measurements of the masses and radii of hot subdwarfs \citep{Charpinet:2008,van-Grootel:2010}. The composition of the EHB star was 5.2\% H, 94.5\% He, $5\times10^{-3}$\% C, 0.1\% N, and 0.03\% O.  

\subsection{Tidal Disruption Models}
Here we briefly describe our tidal disruption models, and we refer the reader to \citet{Clausen:2011} and \citet{Strubbe:2009} for a complete description.  We assume that the HB star is on a parabolic orbit with $R_{\rm P} \le R_{\rm T} $, where $R_{T}$ is the tidal disruption radius $R_{\rm T} = R_{\star}(M_{\rm BH}/R_{\star})^{1/3}$.  The strength of the encounter is set with the tidal disruption parameter $\beta = R_{\rm T}/R_{\rm P}$.  After tidal disruption, half of the disrupted star falls back to the BH and forms a thin accretion disc.  The rate at which material falls back declines with time as $\dot{M} \propto t^{-5/3}$.  When $\dot{M}$ exceeds the BH's Eddington rate, we assume that the excess is driven away in an outflow but we do not include the outflowing material in our models (see \citealt{Strubbe:2011} for an exploration of the outflow.)  This results in the accretion flare having a constant luminosity of $L = 1.3 \times 10^{38} ~M_{\rm BH}~{\rm erg~s^{-1}}$ during this super-Eddington phase.  

The other half of the debris remains unbound and moves away from the BH with a range of velocities given by $v(\phi) = (2GM_{\rm BH}/R_{\rm P})^{1/2} \cot (\phi/2)$, where $\phi$ is the azimuthal angle in the orbital plane and is in the range $(\pi - (12 R_{\star}/R_{\rm P})^{1/2} ) < \phi < \pi$.  The density in the debris tail is given by $n(\phi,t) = n_{0}[v(\phi)t]^{-3}$, where $t$ is the time since tidal disruption and the normalisation factor $n_{0}$ is set such that the total mass in the debris tail is $M_{\star}/2$.  We divide the debris tail into six azimuthal zones and compute photoionisation models for each zone with version 10.00 of {\em Cloudy} \citep{Ferland:1998}, using the density determined by the dynamical model and the composition given by the \mesa models.  The spectral energy distribution (SED) and intensity of the ionising radiation are determined from the dynamical model of the bound debris. We assume a standard $\alpha$-disc model with $\alpha = 0.1$, an inner radius of 3 Schwarzschild radii, an outer radius of $2\; R_p$, and mass accretion rate equal to the mass fallback rate described above. Using the range of temperatures in the accretion disc, we construct a multicolour blackbody which we us as the ionising continuum for the {\em Cloudy} models.  We have not added an X-ray power law to the SEDs used in these models.  Photoionisation by X-rays can result in significant heating of the unbound material and affect the luminosities of collisionally excited emission lines.  However, \citet{Clausen:2011} found that the luminosities of the optical emission lines of interest here were not sensitive to this additional component.

We modelled the emission lines produced with 20 different sets of initial conditions to explore the parameter space in BH mass ($30~M_{\sun}\le M_{\rm BH} \le10^{4} M_{\sun}$), HB star structure and composition (red clump and EHB), and tidal disruption parameter $\beta$ ($0.1\le \beta \le1$). 
     
\section{Results}

\label{results}
Each combination of $M_{\rm BH}$, HB star type, and $\beta$ that we considered predicted \nii~and \oiii~emission lines with luminosities $\sim 10^{36}~{\rm erg~s^{-1}}$ at some point during the post tidal disruption evolution.  However, only those with a red clump HB star, a BH with $M_{\rm BH} \la 200~M_{\sun}$, and $\beta \sim 1$ were able to reproduce the emission line luminosity ratios reported in I10.  These constraints are \nii/H$\alpha > 7$, \oiii/H$\beta > 5$, and \nii/\oiii $\sim 1$.   We plot the time evolution of these ratios for a subset of our models in \autoref{ratios}.  In the models plotted here, we have used our fiducial red clump HB star and assumed that $R_{\rm P} = R_{\rm T}$, but varied $M_{\rm BH}$. 
\begin{figure}
	\centering
	\includegraphics[width=0.5\textwidth]{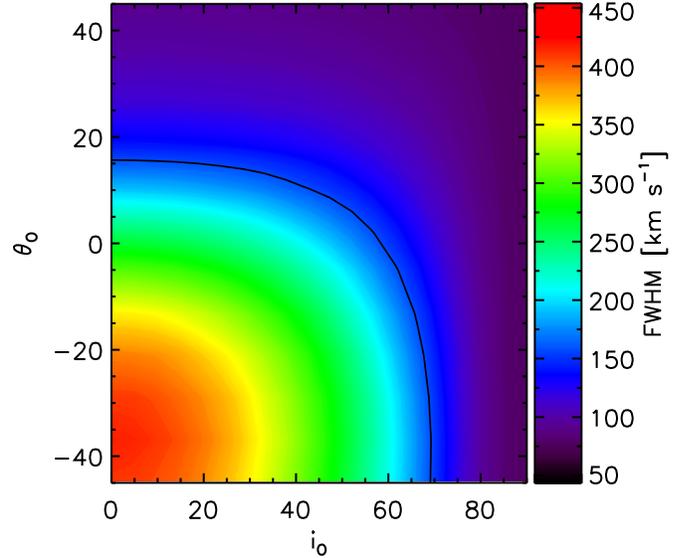}
	\caption{FWHM of the \nii~emission line 125 years after tidal disruption, synthesised from a model using the red clump star, $\beta = 1$, and $M_{\rm BH}= 100~M_{\sun}$. The angles $\theta_{0}$ and $i_{0}$ describe the orientation of the observer (see fig 1 in \citet{Clausen:2011}).  The black curve marks the observed FWHM. Any pair of $\theta_{0}$ and $i_{0}$ along the black line produces a FWHM that is consistent with observations.\label{fwhm}}
\end{figure}
\begin{figure}
	\centering
	\includegraphics[width=0.55\textwidth]{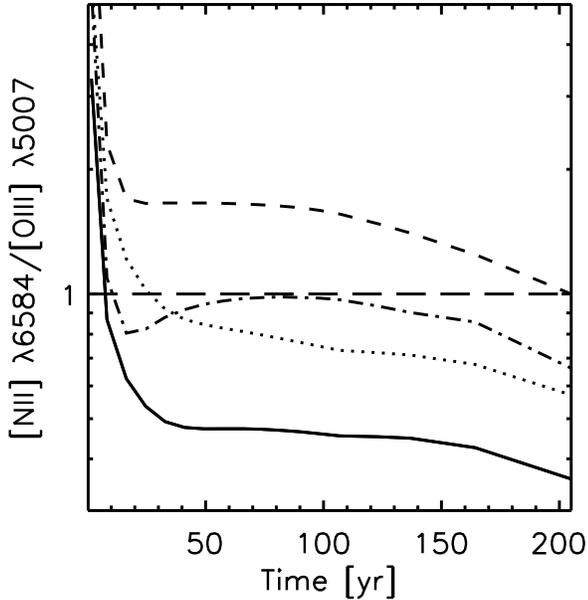}
	\caption{Evolution of \jnii/\joiii~for HB stars of different composition.  We show two sets of models -- young HB stars 30 Myr after He ignition when the nitrogen and oxygen abundance are equal ($M_{\rm BH} = 50 M_{\sun}$ dashed; $M_{\rm BH} = 100 M_{\sun}$ dotted) and the same models but with the HB star older so that 80 Myr of He burning has made oxygen twice as abundant as nitrogen ($M_{\rm BH} = 50 M_{\sun}$ dot-dashed; $M_{\rm BH} = 100  M_{\sun}$ solid).  For a 100 $M_{\sun}$ BH, the observed \jnii/\joiii~ratio cannot be reproduced once helium burning has created an oxygen abundance that is twice the nitrogen abundance.   \label{abds}}
\end{figure}

In each of these models, the Balmer lines outshone the [\ion{O}{3}] and [\ion{N}{2}] lines initially.  Then, as the debris cloud expanded and $n$ decreased, the luminosity of the Balmer lines declined because the volume emissivity of these permitted transitions is proportional to the product of the electron and ionised hydrogen number densities.  At the same time, the \oiii~and \nii~lines became brighter.  These forbidden lines are produced when ions are collisionally excited into a metastable state and then de-excite radiatively.  As the density in the unbound debris decreased, the collision rate declined, allowing a larger fraction of the excited ions to decay radiatively, as opposed to collisionally, and the forbidden emission-lines grew brighter.  As a result, all of the models shown in \autoref{ratios} can satisfy the \joiii/H$\beta$ constraint, provided we are observing the debris more than 40 years after tidal disruption.  The models with $M_{BH} = 10^{3}$ and $10^{4}~M_{\sun}$ were unable to reproduce the observed \jnii/H$\alpha$ ratio.  These more massive BHs permit higher accretion rates that led to an increased flux of ionising photons.  By the time the density in the debris was low enough to produce a significant \jnii~luminosity, \jnii~emission was suppressed because the cloud had become so highly ionised that much of the N$^{+}$ was further ionised to N$^{++}$.   

This over-ionisation of the unbound debris also drove the evolution of the \jnii/\joiii~ratio.  In all of the models, the ratio steadily declined as an increasing portion of the oxygen and nitrogen in the cloud became doubly ionised. With $M_{\rm BH} = 10^{3}$ or $10^{4}~ M_{\sun}$, the unbound debris became too highly ionised for the \jnii/\joiii ratio to be $\sim 1$ within ten years of tidal disruption (see   \autoref{ratios}).  We found that in models with $M_{\rm BH} < 200 M_{\sun}$, the decline is slow enough that the luminosities of the two forbidden emission lines can differ by less than a factor of two for up to 200 years.  

\autoref{odd}, shows a number of alternative models for strong disruption and/or blue HB stars.  These models fail to reproduce the observed \jnii/\joiii~ratio for an extended period of time, in contrast to our models of weakly disrupted red HB stars.  Like the models involving high mass BHs discussed above, the unbound debris cloud becomes too highly ionised to produce significant \jnii~emission.  Using a simple estimate of the ionisation parameter $U=n_{\gamma}/n_{H}$,  we can illustrate why some debris clouds become over ionised and others do not.  Here, $n_{\gamma}$ is the number density of photons with energies above 13.6 eV and $n_{H}$ is the hydrogen number density. Recasting the ionisation parameter in terms of our model parameters, we find $U \propto R_{\star}^{-1/2} \beta^{3} \dot{M}^{3/4} t$.   Debris from stars with a larger kinetic energy at pericentre passage, due to a more compact star with smaller $R_{T}$, an orbit with $\beta > 1$, or a more massive BH, spanned a wider range in velocity and quickly expanded to a low density, resulting in a large $U$. As mentioned above, more massive BHs lead to a higher flux of ionising photons, also contributing to increased $U$.  During the super-Eddington phase, $\dot{M} \propto M_{\rm BH}$ was constant and the ionisation parameter increased with time.  The N$^{+}$ fraction reached a maximum\footnotemark[1] when $U \sim 0.06$, and at larger values of $U$, the luminosity of the \jnii~line declined.  Keeping the ionisation state in the unbound debris low enough to produce significant \jnii~emission after the super-Eddington phase had ended required the relatively large radius of the red clump star and a BH of relatively low mass, $M_{\rm BH} < 200~M_{\sun}$.  After the super-Eddington phase, $U$ declined as $t^{-1/4}$ and N$^{++}$ recombined into N$^{+}$.  This drove the \jnii/\joiii~ratio back towards unity in some cases (see models with $\beta > 1$ in \autoref{odd}).  However, at this point the density in the cloud is so low that the luminosities of the lines are well below the observed values.         
\footnotetext[1]{The value of $U$ at which this maximum occurs is abundance dependent. The value given is for the red clump star composition. The maximum occurs at a much smaller value of $U$ if solar abundances are used.}  

Finally, by combining the dynamical model with the photoionisation model, we synthesised the emission line profiles of the \jnii~and \joiii~lines.  \autoref{fwhm} shows the range of observed full width at half maximum (FWHM) of the \jnii~line produced in a the model of a red clump star on an orbit with $\beta = 1$ being tidally disrupted by a $100 M_{\sun}$ BH, 125 years after tidal disruption.  The angles $\theta_{0}$ and $i_{0}$ describe the orientation of the observer, and are the angle between the line of site and the direction of pericentre and the orbital plane, respectively.  The black line shows the observed FWHM $= 140 {\rm~km~s^{-1}}$.  For a range of observer orientations, the FWHM of the synthesised emission lines is consistent with the observations.       

\begin{figure}
	\centering
	\includegraphics[width=0.55\textwidth]{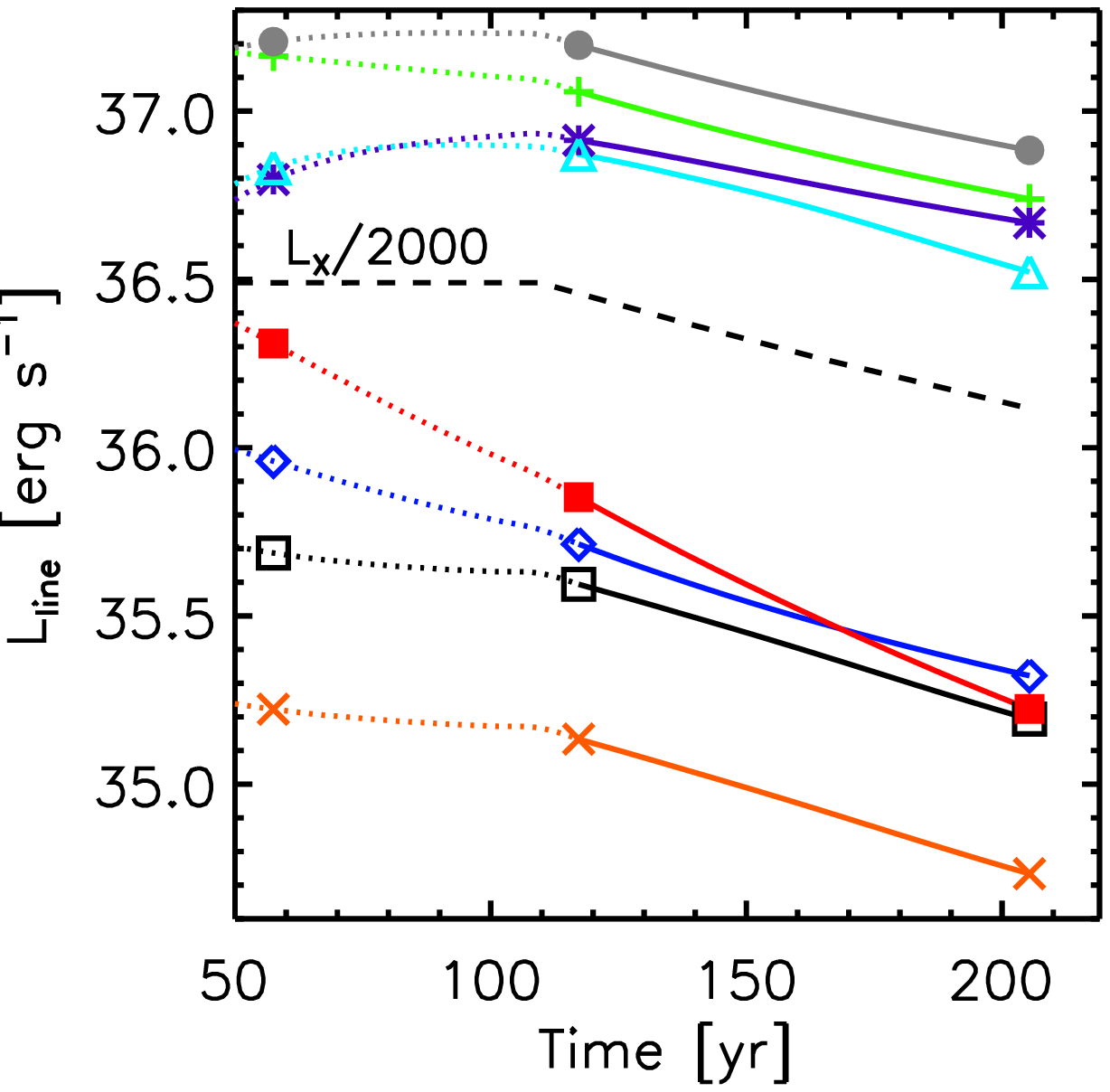}
	\caption{ Light curves from a model using the fiducial red clump HB star, $\beta=1$, and $M_{\rm BH} = 100~M_{\sun}$.  Emission line light curves are shown for  \ion{N}{5} $\lambda1240$ (grey, filled circle), \ion{C}{4} $\lambda1548$ (green, plus), \oiii~(purple, asterisk), \nii~(light blue, triangle), \ion{N}{2} $\lambda5755$ (red, filled box), \ion{O}{3} $\lambda4363$ (dark blue, diamond), H$\alpha$ (black, open box), H$\beta$ (orange, x).  The dashed curve shows $L_{\rm X}$ scaled down by a factor of 2000. Note the bright \ion{N}{5} and \ion{C}{4} doublets predicted by the model. The dotted portion of each curve shows the line luminosity while the fallback rate is super-Eddington and our model is uncertain.\label{lightcurve}}
\end{figure}
                
\section{Discussion}
\label{discussion}
We have shown that the photoionised debris of a tidally disrupted HB star can account for the emission lines observed in optical spectra presented in I10.  Reproducing the line ratios within the context of this model requires $M_{\rm BH} \la 200~M_{\sun}$.  In these scenarios, the super-Eddington phase lasts between 90 and 155 years.  Since we have not included the outflow produced during this phase in our models, the emission line luminosities calculated during this period are uncertain. However, the modelled emission line luminosities remain consistent with the observations after the super-Eddington phase has ended.  Furthermore, the X-ray luminosity of the accretion flare does not decline to the observed value of $L_{\rm X} \sim 2 \times 10^{39}~{\rm erg~s^{-1}}$ until after the mass fall back rate has dropped below the Eddington rate.  Archival ROSAT HRI images establish that the X-ray source has been on since at least Jan. 1996.  At times greater than $t=200$ years after disruption, our assumption of steady state accretion begins to break down because the viscous time in the accretion disc approaches $t$.  We have also checked that the conditions in the photoionised debris change slowly enough that the unbound debris remained in photoionisation equilibrium for the entire period explored in our models.

We did not consider spinning BHs in this study.  The last stable orbit around a spinning BH is closer to the event horizon, allowing the accretion disc to extend to smaller radii than the discs considered here.  Accretion discs around spinning BHs produce a larger flux of ionising photons with an SED that peaks at a higher photon energy because the maximum temperature in these discs is larger than that of accretion discs around non-spinning BHs.  However, unlike the supermassive BHs found in the cores of galaxies, BHs in  GCs have not gained an appreciable portion of their mass though disc accretion.  \citet{Belczynski:2008} found that any accretion that does occur while the BH is in a mass transferring binary does not substantially increase its spin.   However, it is possible that the GC BHs are born with a large spin.  If this is the case, the HB tidal disruption scenario remains a plausible explanation for the observed emission lines, but the BH mass inferred from the models would change by a factor of a few.         

The \jnii/\joiii~ratio is sensitive to the abundance of nitrogen and oxygen in the HB star.  In the models discussed above, we used the composition of the HB star at the onset of helium burning, when the ratio of nitrogen to oxygen is largest.  Helium burning destroys nitrogen and creates oxygen, so N/O declines as the star evolves on the HB.  \autoref{abds} shows how \jnii/\joiii~responds to changes in composition of the HB star.  In these models, we used the composition of the red clump star at later times, as computed in the \mesa models, corresponding to points where the nitrogen and oxygen abundance are equal, and where the nitrogen abundance is half that of oxygen.  The mass and radius of the star did not change appreciably in this time, so we used the same values as in our fiducial model; and we used $\beta = 1$.  When $M_{\rm BH} = 50~M_{\sun}$, the models with both abundance sets are consistent with the observed \jnii/\joiii.  The \jnii/\joiii~ratio drops too quickly to be consistent with observations when $M_{\rm BH} = 100~M_{\sun}$ and the oxygen abundance is twice that of nitrogen.  This sets the period during which the HB star can be disrupted by the BH and produce the observed emission line spectrum.  For the $100~M_{\sun}$ BH, the HB star must be disrupted within 30 Myr of helium ignition to satisfy the \jnii/\joiii~constraint, while a $50~M_{\sun}$ BH can disrupt the HB star within 80 Myr of helium ignition and meet the constraint.   

It is appealing that the tidal disruption models require a red horizontal branch star to meet the observational constraints on the emission line ratios.  The host globular cluster is red and likely exhibits the red horizontal branch morphology typical of a metal rich cluster.  However, given the predicted rate at which GC BHs tidally disrupt stars, it is surprising that such an event might have been observed.  Employing the tidal disruption rate equation computed by \citet{Baumgardt:2004}, with the mass and radius of a red clump HB star, $M_{\rm BH} = 100~M_{\sun}$, a core density of $5 \times 10^5~{\rm pc^{-3}}$, a core velocity of $10~{\rm km ~s^{-1}}$, and assuming that 5\% of the stars in a globular cluster's core are HB stars, we find a HB tidal disruption rate of $1.2 \times 10^{-10}~{\rm yr^{-1}~GC^{-1}}$.  Given that the bright \jnii~and \joiii~emission lines persist for 200 years, and that the  space density of globular clusters is $\sim 6~{\rm Mpc^{-3}}$ \citep{Brodie:2006,Croton:2005}, at any time there should only be $5\times 10^{-3}$ observable HB tidal disruptions within the 20 Mpc distance to NGC 1399. Furthermore, if the NGC 1399 GC source is the aftermath of a HB tidal disruption, it is also puzzling that several main sequence tidal disruptions have not been observed because they because they should occur more frequently \citep{Baumgardt:2004}.

The low disruption rate and the dearth of main sequence tidal disruptions are only an issue if the tidal disruption occurs through the conventional loss cone orbit channel.  One possibility that would result in the preferential disruption of post main sequence stars is for the disrupted star to be bound to the BH.  If this is the case, tidal disruption could be triggered in two different scenarios; 1) a scattering induced merger or 2) a triple decay.  In a scattering induced merger, the HB star progenitor is bound to the BH, expands to fill its Roche Lobe while on the giant branch, and transfers material to the BH.  During mass transfer, the semi-major axis of the binary expands, greatly increasing the likelihood that the binary will undergo an encounter with another star in the GC. Such an encounter could impulsively increase the eccentricity and decrease the semi-major axis of the binary and result in the tidal disruption of the bound star \citep{Davies:1995,Sigurdsson:1993a,Sigurdsson:1995}.   Successive distant encounters can also to drive the eccentricity high enough for $R_{\rm P} = R_{\rm T}$.  We have estimated the timescale for this process using the expression given in \citet{Maccarone:2005}, which uses the encounter cross sections computed by \citet{Heggie:1996}.  In a binary with a post-mass-transfer semi-major axis of 1 AU containing a  $100~M_{\sun}$ BH, a $0.65~M_{\sun}$ HB star will be driven to the tidal disruption radius in 10 Myr.  This is well within the 30 Myr window required to produce \jnii~and \joiii~emission lines of similar luminosity.  Here we have assumed that the field stars have mass 0.8 $M_{\sun}$, a cluster density of $5 \times 10^5~{\rm pc^{-3}}$, and a velocity of dispersion of $10~{\rm km ~s^{-1}}$.  A scattering induced merger would be more likely in a dense GC.

In the triple decay, the BH and HB star progenitor are initially the inner binary in a stable, hierarchal triple system.  Again, evolution of the HB progenitor results in a phase of mass transfer and expansion of the inner binary's semi-major axis.  In this case, expansion of the inner binary drives the triple to instability and triggers the tidal disruption.  A similar scenario was discussed by \citet{Perets:2012}, only these authors invoked mass loss, not mass transfer, to drive the expansion of the inner binary. The triple decay scenario would require a medium density GC, because triple systems are unstable to external perturbations in dense GCs.  Both the scattering induced merger and triple decay scenarios increase the odds that tidal disruptions are linked to post main sequence stars, but determining if disruption through either scenario would occur more frequently than the loss cone tidal disruptions discussed above is beyond the scope of this paper.
                                     
Our models also predict several strong UV emission lines.  These lines could be used to test the tidally disrupted HB star hypothesis for the ULX in NGC 1399 and to further constrain the mass of the BH.  The strongest lines are \ion{N}{5} $\lambda\lambda 1239,1243$, \ion{O}{6} $\lambda\lambda{1032,1038}$, \ion{N}{4}] $\lambda 1486$, and \ion{C}{4} $\lambda\lambda 1548,1550$.  These lines should have FWHM similar to the observed optical lines.  The bright \ion{C}{4} doublet could be used to constrain the composition of the disrupted star and narrow the allowed range in $M_{\rm BH}$.  The light curves for the \ion{N}{5} and \ion{C}{4} emission lines are shown in \autoref{lightcurve} along with the emission line light curves for the optical lines with observational constraints.  In contrast, \citet{Maccarone:2011} do not report any strong UV lines from their models of photoionised  RCB star winds.   

\section{Conclusion}
We have presented models for the emission lines generated in the photoionised debris of tidally disrupted HB stars.  Within the assumptions made in our models, the emission line spectra produced when a red clump star is disrupted by a BH with $M_{BH} \la 200$ are consistent with the optical emission lines observed by I10 in the ULX hosting globular cluster in NGC 1399.  The models also make testable predictions about the source's UV-emission line spectrum. Observations of the UV-emission lines could significantly improve our understanding of this source and are likely possible with the {\em Hubble Space Telescope}.  
\vspace{0.5cm}

{\bf \noindent ACKNOWLEDGMENTS}

\noindent We thank the anonymous referee for a prompt and helpful review.  DC is supported by the PSU Academic Computing Fellowship.  The authors thank the Aspen Center for Physics for its hospitality.     

 \bibliography{hb_unbound.bib}
 
\end{document}